\documentclass[11pt]{article}
\usepackage[journal=jquant,stage=submission,logging=stdout]{jsty3-author}

\usepackage{ifthen,amsthm,amsmath,amsxtra,amsfonts,dsfont,graphicx,bm,tikz,scalerel,wasysym,physics,bbm,graphicx,mathtools,tensor,mathrsfs}

\usepackage{CircuitTikz}

\newcommand{\unit}{\mathbbm{1}}

\theoremstyle{plain}

\title{Exact subsystem dynamics in the deterministic Floquet-PXP model}

\author{1}{Katja Klobas}{k.klobas@bham.ac.uk}{0000-0001-6791-8594}
\affiliation{1}{School of Physics and Astronomy, University of Birmingham, Edgbaston, Birmingham, B15 2TT, UK}
\abstract{The dynamics of local subsystems in a thermodynamically large quantum many-body system can be understood as effectively open as the system produces its own effective bath. The action of this bath can be characterised in terms of the so-called influence matrices. In generic situations, the complexity of these objects grows unfavourably with time, however, there exist solvable cases where influence matrices can be characterised exactly even in the presence of non-trivial interactions. Here we show that Rule 201, a deterministic version of the Floquet-PXP model, is one of these solvable instances. Indeed, it admits influence matrices given by a finite-dimensional matrix-product operator (MPO) that solves a finite set of algebraic conditions. We provide the solution, and characterise multi-time autocorrelation functions.}

\begin{document}
\section{Introduction}\label{sec:intro}
Integrable models form the backbone of our understanding of equilibrium
thermodynamics in the presence of
interactions~\cite{takahashi1999thermodynamics,yang1969thermodynamics,korepin1993quantum}.
They allow for an exact description of equilibrium properties, in both ground
states and thermal states, as well as the quasi-stationary behaviour observed
at late times after quantum
quenches~\cite{calabrese2016introduction,essler2016quench,caux2016quench,doyon2020lecture,alba2021generalizedhydrodynamic,bastianello2022introduction}.
When one is interested in truly out-of-equilibrium properties, however,
integrable models are far harder to analyse. It is therefore desirable to find
a new class of interacting systems whose dynamics can be exactly described as
well.

Recently, it was understood that a convenient setting to search for such models
is that of \emph{quantum circuits}, systems defined in discrete space-time with
time evolution given as a sequence of discrete time-steps consisting of local
updates --- gates --- which act nontrivially only on a small subset of degrees
of freedom~\cite{bertini2026quantum}. Quantum circuits arise as a convenient
trick to approximate numerically the time-evolution of autonomous systems, and as
such can exhibit the same dynamical features. A more modern view, however, is to
considered as bona fide dynamical systems in their own right. Exact solutions in quantum
circuits can be generally achieved in two ways, either through disorder
averaging~\cite{potter2022entanglement,fisher2023random}, or by choosing the
gates so that they satisfy suitable algebraic
relations~\cite{bertini2026exactly}. Both these approaches have been very
fruitful and have aided our understanding of non-equilibrium phenomena such as
information scrambling, dynamics of entanglement, and the onset of chaos.

A prominent example of quantum circuits that are solvable without the need to introduce disorder are \emph{dual-unitary circuits}~\cite{bertini2019exact}, which consist of gates generating unitary dynamics also when the roles of space and time are exchanged. This restriction turned out to allow exact and explicit calculations of several dynamical and spectral properties~\cite{bertini2026exactly}, however, the solvability condition also limits the available phenomenology. For instance, non-vanishing dynamical correlation functions only exist on the edges of the causal light-cone. This has motivated a search for alternative, less restrictive, conditions that are still based upon the idea of exchanging the roles of space and time.
In the context of the dynamics of local observables, this idea
takes the form of a \emph{transverse contraction} of the tensor
network, as introduced in Ref.~\cite{banuls2009matrix} (see
also~\cite{muller2012tensor,hastings2015connecting,frias2022lightcone}).
Namely, the tensor network describing an expectation value of a local
observable can be equivalently understood as resulting from the evolution in
the space direction, given by the \emph{space transfer matrix}. This point of
view is convenient when considering dynamics of strictly local subsystems, as
in the thermodynamic limit the effect of the rest of the system is completely
encoded in the leading eigenvectors (also referred to as \emph{fixed points})
of the transfer matrix. Ref.~\cite{lerose2021influence} proposed that these
eigenvectors are not just a convenient technical tool to contract certain
tensor networks, but can be understood as the \emph{effective baths} that
induce the thermalization of finite subsystems in the thermodynamically large
system.  In analogy with the Feynman-Vernon influence
functional~\cite{feynman1963theory} a fixed point is also referred to as an
\emph{influence matrix}~\cite{lerose2021influence,sonner2021influence}. By
construction influence matrices give access to physics of local observables
in various dynamical
regimes~\cite{piroli2020exact,lerose2021influence,klobas2021exact,klobas2021exactrelaxation,sonner2022characterizing,thoenniss2023nonequilibrium},
and, furthermore, they can be used to study the entanglement growth after a
global quench~\cite{bertini2019entanglement,piroli2020exact,klobas2021entanglement,ippoliti2022fractal,bertini2022growth,klobas2024nonequilibrium}.

In dual unitary circuits evolving from compatible initial states, influence
matrices factorise into featureless maximum-entropy
states~\cite{piroli2020exact}, and so the effective bath acting on the
subsystem is perfectly Markovian. This realisation suggests a natural avenue
for solvability beyond dual unitarity by characterising instances where
influence matrices can be expressed exactly and yet they retain some nontrivial
information in
time~\cite{klobas2021exact,klobas2021exactrelaxation,klobas2021entanglement,giudice2022temporal,yu2024hierarchical,foligno2024quantum,wang2024exact,klobas2024exact,defazio2024exact,hubner2026influence,pickering2026asymptotically,rampp2026infinite}.
A prominent example is that of the reversible cellular automaton referred to as
\emph{Rule 54}~\cite{bobenko1993two,buca2021rule}, which for compatible initial
states admits a constant-Schmidt-rank representation of influence
matrices~\cite{klobas2021exact,klobas2021exactrelaxation}. The exact expression
for fixed points follows from a set of local algebraic relations fulfilled by a
finite set of operators, which can be fulfilled by a set of $3$-dimensional
matrices.

A natural question is whether Rule 54 is the only one of its kind, or similar
structure can also be found in other systems.  In this paper we show that the
same set of algebraic relations can indeed be solved for different choices of
local gates, by providing an example of the Floquet-PXP cellular automaton,
also referred to as \emph{Rule 201}, introduced
in~\cite{wilkinson2020exact,iadecola2020nonergodic}. The model can be understood as
a deterministic point of the Floquet version of the PXP
model~\cite{lesanovsky2011many,turner2018weak,giudici2024unraveling}, and
has recently been also experimentally
implemented~\cite{white2026quantum,cesa2026engineering}. We construct the
influence matrices corresponding to Gibbs states, which take form of the MPO
with bond dimension 12. This allows us to completely characterise one-site
multi-time correlation functions and gives insight into potential Bethe
equations for this model.

The rest of the manuscript is organised as follows. In section~\ref{sec:setup} we
discuss the setting and define the dynamics, and then in section~\ref{sec:infmat}
we discuss the stationary states and give the form of the influence matrices.
In section~\ref{sec:MultiTimeDiag} we show that the obtained fixed points immediately
provide a convenient reformulation of multi-time correlation functions and we 
characterise their decay. Finally, in section~\ref{sec:conclusions} we conclude with
some final remarks.

\section{The setup}\label{sec:setup}
The model is defined on a (periodic) qubit chain of length $L$, with the dynamics
given in terms of two distinct steps,
\begin{equation}
  \ket{\psi(t+1)} =
  \begin{cases}
    \mathbb{U}^{\mathrm{e}} \ket{\psi(t)},\qquad  &t\equiv 0\pmod{2},\\
    \mathbb{U}^{\mathrm{o}} \ket{\psi(t)},\qquad  &t\equiv 1\pmod{2},
  \end{cases}
\end{equation}
where $\ket{\psi(t)}\in\left(\mathbb{C}^2\right)^{\otimes L}$ and 
$\mathbb{U}^{\mathrm{e/o}}$ are one time-step time evolution operators
corresponding to even and odd time steps. They are both expressed as 
products of mutually commuting local three-site operators,
\begin{equation}
  \mathbb{U}^{\mathrm{e}} = \prod_{j=1}^{L/2} U_{2j-1,2j,2j+1},\qquad
  \mathbb{U}^{\mathrm{o}} = \prod_{j=1}^{L/2} U_{2j,2j+1,2j+2},
\end{equation}
where $U_{j-1,j,j+1}$ acts nontrivially on the triplet of sites $(j-1,j,j+1)$,
\begin{equation}
  U_{j-1,j,j+1} = \unit^{\otimes j-2}\otimes U \otimes \unit^{\otimes L-j-1},
\end{equation}
and $U$ is a $8\times 8$ matrix that leaves the left and right sites intact, while
the central qubit is changed according to the three-site deterministic rule
$\chi:\mathbb{Z}_2\otimes\mathbb{Z}_2\otimes\mathbb{Z}_2\to\mathbb{Z}$,
\begin{equation}
  U^{s_1^{\prime} s_2^{\prime} s_3^{\prime}}_{s_1^{\phantom{\prime}} s_2^{\phantom{\prime}} s_3^{\phantom{\prime}}} =
  \delta_{s_1^{\prime},s_1^{\phantom{\prime}}}
  \delta_{s_2^{\prime}, \chi(s_1,s_2,s_3)}
  \delta_{s_3^{\prime},s_3^{\phantom{\prime}}},
\end{equation}
with
\begin{equation}
  \chi(s_1,s_2,s_3)=(1-2s_2)\delta_{s_1,0}\delta_{s_3,0}+s_2.
\end{equation}
Note that $U$ can be equivalently expressed as
\begin{equation}
  U=P_{0}\otimes \left(\sigma^x-\unit\right)  \otimes P_0 + \unit\otimes\unit\otimes \unit,
\end{equation}
where $P_0=\left(\sigma^z+\unit\right)/2$, $\sigma^{x,y,z}$ are Pauli matrices and $\unit$ is a
$2\times 2$ identity matrix.

\begin{figure}
  \centering
  \includegraphics[width=0.65\textwidth]{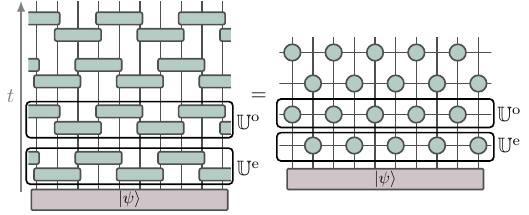}
  \caption{\label{fig:brickwork} Diagrammatic representation of time-evolution of an initial
  state $\ket{\psi}$.}
\end{figure}

Introducing the following symbol for the local $3$-site gate,
\begin{equation}
  U= \begin{tikzpicture} [baseline={([yshift=-0.6ex]current bounding box.center)},scale=0.35]
    \foreach \x in {1,...,3}{\tgridLine{\x}{-1}{\x}{1}}
    \prop{1}{3}{0}{colU}
  \end{tikzpicture},
\end{equation}
time evolution can be represented as a brickwork-like quantum circuit shown on the l.h.s\ of  
figure~\ref{fig:brickwork}. However, since all the local gates applied in the same time-step
commute, this is not the most convenient graphical representation, as the illustration does
not share the same symmetry. Rather than that, it makes more sense to think of each time-step
as a matrix-product operator (MPO), by introducing the tensor that encodes the time-evolution
rule $\chi$ and is given by the following matrix elements,
\begin{equation}\label{eq:defBigT}
  \begin{tikzpicture} [baseline={([yshift=-0.6ex]current bounding box.center)},scale=0.35]
    \tgridLine{0}{-1}{0}{1}
    \tgridLine{-1}{0}{1}{0}
    \node at (-1.5,0) {\scalebox{0.8}{$s_1$}};
    \node at (0,-1.5) {\scalebox{0.8}{$s_2$}};
    \node at (1.5,0) {\scalebox{0.8}{$s_3$}};
    \node at (0,1.5) {\scalebox{0.8}{$s_4$}};
    \bCircle{0}{0}{colU}
  \end{tikzpicture}=\delta_{s_4,\chi(s_1,s_2,s_3)},
\end{equation}
while the intersecting lines imply that all the incoming legs are in the same state,
\begin{equation}\label{eq:defSmallT}
  \begin{tikzpicture} [baseline={([yshift=-0.6ex]current bounding box.center)},scale=0.35]
    \tgridLine{0}{-1}{0}{1}
    \tgridLine{-1}{0}{1}{0}
    \node at (-1.5,0) {\scalebox{0.8}{$s_1$}};
    \node at (0,-1.5) {\scalebox{0.8}{$s_2$}};
    \node at (1.5,0) {\scalebox{0.8}{$s_3$}};
    \node at (0,1.5) {\scalebox{0.8}{$s_4$}};
  \end{tikzpicture}=\delta_{s_1,s_2}\delta_{s_2,s_3}\delta_{s_3,s_4}.
\end{equation}
Using definitions~\eqref{eq:defBigT} and~\eqref{eq:defSmallT}, the time-evolution can be
equivalently represented as a staggered MPO graphically given by the r.h.s.\ of
figure~\ref{fig:brickwork}.

\begin{figure}
  \centering
  \includegraphics[width=\textwidth]{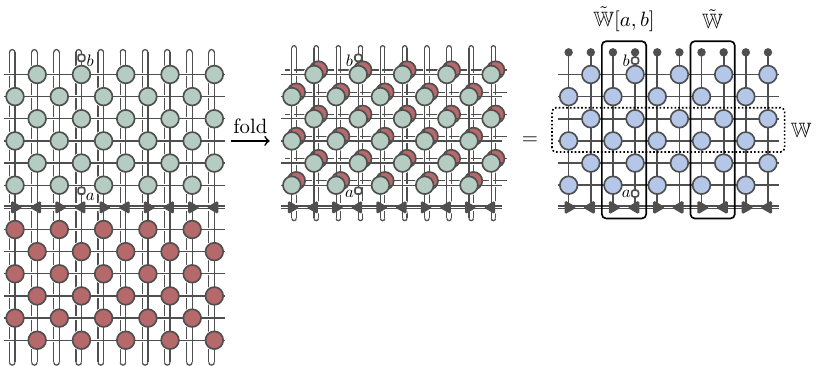}
  \caption{\label{fig:folding} Graphical representation of a two-point dynamical correlation function on a Gibbs state at the same position.}
\end{figure}

Let us now consider a dynamical correlation function on a stationary state between two local observables at the same position,
\begin{equation}\label{eq:defCab}
  C_{a,b}(t)
  =\frac{1}{Z}\tr\left[
    \left(\mathbb{U}^{\rm o}\mathbb{U}^{\rm e}\right)^{\dagger\, t}
    \rho \, a
    \left(\mathbb{U}^{\rm o}\mathbb{U}^{\rm e}\right)^{t} b \right],
\end{equation}
where $a$ and $b$ are local Hermitian operators, and $Z$ is given by the normalisation of the stationary state,
\begin{equation}
  Z=\tr[\rho].
\end{equation}
The correlation function is for the example of two one-site observables $a$ and $b$ shown diagrammatically on the l.h.s.\ of figure~\ref{fig:folding}, where we also assume that $\rho$ takes a staggered matrix-product-operator (MPO) representation (staggered triangles), and the top bottom half of the tensor network corresponds to the Hermitian adjoint of the top half. Since the two halves are coupled together it is convenient to imagine to bend the bottom part behind the top one, and define a new set of tensors acting on the two copies of the Hilbert space together,
\begin{equation}
  \begin{gathered}
  \begin{tikzpicture}[baseline={([yshift=-0.6ex]current bounding box.center)},scale=0.375]
    \gridLine{0}{1}{0}{-0.5}
  \end{tikzpicture}=
  \begin{tikzpicture}[baseline={([yshift=-0.6ex]current bounding box.center)},scale=0.375]
    \begin{scope}[shift={(0.2,0.2)}]
      \tgridLine{0}{1}{0}{-0.5}
    \end{scope}
    \tgridLine{0}{1}{0}{-0.5}
  \end{tikzpicture},\qquad
  \begin{tikzpicture}[baseline={([yshift=-0.6ex]current bounding box.center)},scale=0.375]
    \gridLine{0}{1}{0}{-0.5}
    \ME{0}{1}
  \end{tikzpicture}=
  \begin{tikzpicture}[baseline={([yshift=-0.6ex]current bounding box.center)},scale=0.375]
    \begin{scope}[shift={(0.2,0.2)}]
      \tgridLine{0}{1}{0}{-0.5}
    \end{scope}
    \tgridLine{0}{1.2}{0}{-0.5}
    \ttopHook{0.2}{1.2}{0.1}
  \end{tikzpicture},\qquad
  \begin{tikzpicture}[baseline={([yshift=-0.6ex]current bounding box.center)},scale=0.375]
    \gridLine{-1}{0}{1}{0}
    \gridLine{0}{-1}{0}{1}
    \bCircle{0}{0}{FcolU}
  \end{tikzpicture}=
  \begin{tikzpicture}[baseline={([yshift=-0.6ex]current bounding box.center)},scale=0.375]
    \begin{scope}[shift={(0.2,0.2)}]
      \tgridLine{-1}{0}{1}{0}
      \tgridLine{0}{-1}{0}{1}
      \bCircle{0}{0}{colUc}
    \end{scope}
    \tgridLine{-1}{0}{1}{0}
    \tgridLine{0}{-1}{0}{1}
    \bCircle{0}{0}{colU}
  \end{tikzpicture}.
  \end{gathered}
\end{equation}
Using this folding transformation the tensor network can be equivalently represented
by a half smaller one, shown in the r.h.s.\ of figure~\ref{fig:folding}, while at the same time
the local degrees of freedom double: a line in the folded network represents a $2$-qubit state.

The tensor network represented in figure~\ref{fig:folding} is built by repeated
application of the folded time-evolution map $\mathbb{W}$ (see the top-right
picture in figure~\ref{fig:folding}).  However, the same tensor network can be
also interpreted as a repeated application of the \emph{transverse transfer
matrix} (also referred to as \emph{space} or \emph{dual} transfer matrix)
$\tilde{\mathbb{W}}$, defined as two columns of the time-evolution tensors (see
figure~\ref{fig:folding}). The correlation function~\eqref{eq:defCab} can be
easily expressed in terms of the dual transfer matrix as
\begin{equation}
  C_{a,b}(t)=
  \tr \big(\tilde{\mathbb{W}}[a,b]\tilde{\mathbb{W}}^{\frac{L}{2}-1}\big),
\end{equation}
which immediately implies that for large $L$ the dynamics of local observables is characterized
by spectral properties of the transverse map $\tilde{\mathbb{W}}$. In particular, under mild assumptions~\footnote{Specifically, we assume that the MPO of $\rho$ is injective.}, one can show that $\tilde{\mathbb{W}}$ has a dominant and isolated eigenvalue $\Lambda$ satisfying
\begin{equation}
  \lim_{L\to \infty}\frac{\Lambda^{L/2}}{Z}=1.
\end{equation}
If we then denote the corresponding left and right eigenvectors --- referred to also as \emph{influence matrices} --- by $\bra{L}$ and $\ket{R}$,
\begin{equation}
  \bra{L}\tilde{\mathbb{W}}=\Lambda \bra{L},\quad
  \tilde{\mathbb{W}}\ket{R}=\Lambda \ket{R},\quad \braket{L}{R}=1,
\end{equation}
we can expres the thermodynamic limit of the correlation function as
\begin{equation}
  \lim_{L\to\infty} C_{a,b}(t)=\frac{\mel{L}{\tilde{\mathbb{W}}[a,b]}{R}}{\Lambda}.
\end{equation}
This is a very general statement and requires only that stationary state admits
an efficient MPO representation, which is not uncommon for Floquet systems
which tend to quickly heat up to high temperatures.
However, in general the computational complexity of fixed points $\bra{L}$
and $\ket{R}$ grows exponentially fast with time $t$, which limits the practical usefulness
of this representation~\cite{foligno2023temporal,frias2022light}. Nonetheless,
as we will see later, in our case we are able to find efficient representations
of $\bra{L}$ and $\ket{R}$ for a family of \emph{Gibbs states}.

\section{Influence matrices from Gibbs states}\label{sec:infmat}
As was argued by Ref.~\cite{wilkinson2020exact} Rule 201 appears to be integrable whenever it is restricted to the subspace without pairs of neighbouring $1$ configurations. In this case it exhibits three distinct vacuum-like configurations that get mapped into each other under time-evolution, and the domain walls between regions with different vacua act as left and right-moving quasiparticles undergoing nontrivial pair-wise scattering. For the present discussion it is important to recall that 
a Gibbs-like state 
\begin{equation} \label{eq:defGibbs}
  \rho\propto \mathrm{e}^{-\mu_{+} N_{+}-\mu_{-} N_{-}},
\end{equation}
where $N_{\pm}$ are the numbers of the two types of quasi-particles, and
$\mu_{\pm}$ are the corresponding chemical potentials, admits
a simple staggered MPO representation~\cite{wilkinson2020exact},
\begin{equation}
  \rho=
  \begin{tikzpicture}[baseline={([yshift=-0.6ex]current bounding box.center)},scale=0.35]
    \vLine{0.25}{0}{6.75}{0}
    \vleftHook{0.25}{0}
    \vrightHook{6.75}{0}
    \foreach \x in {1,...,6}{\tgridLine{\x}{-0.75}{\x}{0.75}}
    \foreach \x in {1,3,...,6}{\vmpsW{\x}{0}{colLines}}
    \foreach \x in {2,4,...,6}{\vmpsV{\x}{0}{colLines}}
  \end{tikzpicture}
  =
  \smashoperator{\sum_{\substack{s_1,s_2,\ldots,s_L\\ b_1,b_2,\ldots,b_L}}}
  \tr[W_{s_1,b_1}V_{s_2,b_2}\cdots V_{s_L,b_L}] 
  \ketbra{s_1s_2\cdots s_L}{b_1 \cdots b_L}.
\end{equation}
Here we introduced the graphical representation of the stationary MPO,
\begin{equation}
  W_{s,b}=
  \begin{tikzpicture}[baseline={([yshift=-0.6ex]current bounding box.center)},scale=0.35]
    \vLine{-0.75}{0}{0.75}{0}
    \vmpsW{0}{0}{colLines}
    \tgridLine{0}{-0.75}{0}{0.75}
    \node at (0,1.25) {$s$};
    \node at (0,-1.25) {$b$};
  \end{tikzpicture},\qquad
  V_{s,b}=
  \begin{tikzpicture}[baseline={([yshift=-0.6ex]current bounding box.center)},scale=0.35]
    \vLine{-0.75}{0}{0.75}{0}
    \vmpsV{0}{0}{colLines}
    \tgridLine{0}{-0.75}{0}{0.75}
    \node at (0,1.25) {$s$};
    \node at (0,-1.25) {$b$};
  \end{tikzpicture},
\end{equation}
where the matrices $W_{s,b}$, $V_{s,b}$ are $4$-dimensional,
\begin{equation}
  W_{s,b}=
  \delta_{s,b}
  \begin{bmatrix}
    \delta_{s,0}     & 0            & 0            & \xi \delta_{s,0}    \\
    \xi \delta_{s,1} & 0            & \delta_{s,1} & \omega \delta_{s,1} \\
    0                & \delta_{s,0} & 0            & 0 \\
    0                & 0            & \delta_{s,0} & 0
  \end{bmatrix}=\left.
  V_{s,b}\right|_{\xi \leftrightarrow \omega}.
\end{equation}
and the two parameters $\xi,\omega>0$ are the short-hand for
\begin{equation}\label{eq:defFugacities}
  \xi=\mathrm{e}^{-\mu_{+}},\qquad
  \omega=\mathrm{e}^{-\mu_{-}}.
\end{equation}

The stationary state is not normalised, but we have
\begin{equation}
  \lim_{L\to\infty}\frac{Z}{\Lambda^{L/2}}=1,
\end{equation}
where $\Lambda$ is the (isolated) leading eigenvalue of the MPO matrix $T$
\begin{equation}
  T=\smashoperator{\sum_{s_1,s_2,b_1,b_2}}
  W_{s_1,b_1}V_{s_2,b_2}=
  \begin{bmatrix}
    1 & 0 & \xi & \omega \\
    \xi & 1 & \omega & \xi \omega \\
    \omega & 0 & 1 & \xi \\
    0 & 1 & 0 & 0
  \end{bmatrix},
\end{equation}
and $\Lambda$ is the largest (in magnitude) solution to the following quartic equation  
\begin{equation}
  \Lambda^4 - 3 \Lambda^3 + (3-2\xi\omega) \Lambda^2 - (1-\xi\omega) \Lambda
  + (\xi^2-\omega)(\omega^2-\xi) =0.
\end{equation}

\subsection{Algebraic relations for influence matrices}
We wish to characterise influence matrices, i.e.\ left and right leading eigenvectors of the transfer matrix $\tilde{\mathbb{W}}$ and its staggered counterpart $\tilde{\mathbb{W}}^{\prime}$
defined as
\begin{equation}
  \tilde{\mathbb{W}}=
  \begin{tikzpicture}[baseline={([yshift=-0.6ex]current bounding box.center)},scale=0.35]
    \foreach \t in {1,...,6}{\gridLine{-0.75}{\t}{1.75}{\t}}
    \gridLine{0}{0}{0}{7}
    \gridLine{1}{0}{1}{7}
    \vLine{-0.75}{0}{1.75}{0}
    \vmpsW{0}{0}{colLines}
    \vmpsV{1}{0}{colLines}
    \ME{0}{7}
    \ME{1}{7}
    \foreach \t in {1,3,...,6}{
      \bCircle{0}{\t}{FcolU}
      \bCircle{1}{(\t+1)}{FcolU}
    }
  \end{tikzpicture},\qquad
  \tilde{\mathbb{W}}^{\prime}=
  \begin{tikzpicture}[baseline={([yshift=-0.6ex]current bounding box.center)},scale=0.35]
    \foreach \t in {1,...,6}{\gridLine{-0.75}{\t}{1.75}{\t}}
    \gridLine{0}{0}{0}{7}
    \gridLine{1}{0}{1}{7}
    \vLine{-0.75}{0}{1.75}{0}
    \vmpsV{0}{0}{colLines}
    \vmpsW{1}{0}{colLines}
    \ME{0}{7}
    \ME{1}{7}
    \foreach \t in {1,3,...,6}{
      \bCircle{1}{\t}{FcolU}
      \bCircle{0}{(\t+1)}{FcolU}
    }
  \end{tikzpicture}.
\end{equation}
We start with the left eigenvectors $\bra{L}$ and $\bra{L^{\prime}}$
for which we take an ansatz of the same form as in the case of
RCA54~\cite{klobas2021exact,klobas2021exactrelaxation,klobas2021entanglement},
\begin{equation}
  \bra{L}=
  \begin{tikzpicture}[baseline={([yshift=-0.6ex]current bounding box.center)},scale=0.35]
    \mpsWire{0}{0}{0}{7}
    \vLine{0}{0}{1}{0}
    \foreach \t in {1,...,6}{\gridLine{0}{\t}{1}{\t}}
    \foreach \t in {1,3,...,6}{\mpsB{0}{\t}{colMPS}}
    \foreach \t in {2,4,...,6}{\mpsA{0}{\t}{colMPS}}
    \mpsBvecV{0}{0}{colMPS}
    \mpsBvec{0}{7}
  \end{tikzpicture},\qquad
  \bra{L^{\prime}}=
  \begin{tikzpicture}[baseline={([yshift=-0.6ex]current bounding box.center)},scale=0.35]
    \mpsWire{0}{0}{0}{7}
    \vLine{0}{0}{1}{0}
    \foreach \t in {1,...,6}{\gridLine{0}{\t}{1}{\t}}
    \foreach \t in {1,3,...,6}{\mpsA{0}{\t}{colMPS}}
    \foreach \t in {2,4,...,6}{\mpsB{0}{\t}{colMPS}}
    \mpsBvecW{0}{0}{colMPS}
    \mpsBvec{0}{7}
  \end{tikzpicture},
\end{equation}
where 
$\begin{tikzpicture}[baseline={([yshift=-0.6ex]current bounding box.center)},scale=0.3]
  \mpsWire{0}{-0.625}{0}{0.625}
  \gridLine{0}{0}{0.625}{0}
  \mpsA{0}{0}{colMPS}
\end{tikzpicture}$
and
$\begin{tikzpicture}[baseline={([yshift=-0.6ex]current bounding box.center)},scale=0.3]
  \mpsWire{0}{-0.625}{0}{0.625}
  \gridLine{0}{0}{0.625}{0}
  \mpsB{0}{0}{colMPS}
\end{tikzpicture}$
denote \emph{bulk} tensors whose physical-space (horizontal line) components can be thought
of as matrices in the auxiliary space (vertical line).
The top boundary tensor
$\begin{tikzpicture}[baseline={([yshift=-0.6ex]current bounding box.center)},scale=0.3]
  \mpsWire{0}{0}{0}{0.75}
  \mpsBvec{0}{0.75}
\end{tikzpicture}$
is a vector in the auxiliary space, and the bottom boundary tensors
$\begin{tikzpicture}[baseline={([yshift=-0.6ex]current bounding box.center)},scale=0.3]
  \mpsWire{0}{0}{0}{0.75}
  \vLine{0}{0}{0.75}{0}
  \mpsBvecV{0}{0}{colMPS}
\end{tikzpicture}$,
$\begin{tikzpicture}[baseline={([yshift=-0.6ex]current bounding box.center)},scale=0.3]
  \mpsWire{0}{0}{0}{0.75}
  \vLine{0}{0}{0.75}{0}
  \mpsBvecW{0}{0}{colMPS}
\end{tikzpicture}$
are acting both on the auxiliary space corresponding to the MPO of the stationary state
(horizontal line) and on the fixed-point auxiliary space (vertical line). In analogy with
Refs.~\cite{klobas2021exact,klobas2021exactrelaxation}, we impose a set of local relations that should be satisfied by the fixed-point tensors,
\begin{gather} \label{eq:bottomRelations}
  \begin{tikzpicture}[baseline={([yshift=-0.6ex]current bounding box.center)},scale=0.35]
    \mpsWire{0}{0}{0}{4}
    \vLine{0}{0}{2}{0}
    \foreach \t in {1,...,3}{\gridLine{0}{\t}{2}{\t}}
    \mpsA{0}{3}{colMPS}
    \mpsB{0}{2}{colMPS}
    \mpsA{0}{1}{colMPS}
    \mpsBvecW{0}{0}{colMPS}
    \gridLine{1}{0}{1}{3}
    \bCircle{1}{2}{FcolU}
    \vmpsV{1}{0}{colLines}
  \end{tikzpicture}=
  \frac{\Lambda}{\lambda}
  \begin{tikzpicture}[baseline={([yshift=-0.6ex]current bounding box.center)},scale=0.35]
    \mpsWire{0}{0}{0}{4}
    \vLine{0}{0}{1}{0}
    \foreach \t in {1,...,3}{\gridLine{0}{\t}{1}{\t}}
    \mpsC{0}{2}{3}{colMPS}
    \mpsB{0}{1}{colMPS}
    \mpsBvecV{0}{0}{colMPS}
  \end{tikzpicture},\qquad
  \begin{tikzpicture}[baseline={([yshift=-0.6ex]current bounding box.center)},scale=0.35]
    \mpsWire{0}{0}{0}{3}
    \vLine{0}{0}{2}{0}
    \foreach \t in {1,...,2}{\gridLine{0}{\t}{2}{\t}}
    \mpsA{0}{2}{colMPS}
    \mpsB{0}{1}{colMPS}
    \mpsBvecV{0}{0}{colMPS}
    \gridLine{1}{0}{1}{2}
    \bCircle{1}{1}{FcolU}
    \vmpsW{1}{0}{colLines}
  \end{tikzpicture}=
  \lambda
  \begin{tikzpicture}[baseline={([yshift=-0.6ex]current bounding box.center)},scale=0.35]
    \mpsWire{0}{0}{0}{3}
    \vLine{0}{0}{1}{0}
    \foreach \t in {1,...,2}{\gridLine{0}{\t}{1}{\t}}
    \mpsC{0}{1}{2}{colMPS}
    \mpsBvecW{0}{0}{colMPS}
  \end{tikzpicture},\\ \label{eq:bulkTopRelations}
  \begin{tikzpicture}[baseline={([yshift=-0.6ex]current bounding box.center)},scale=0.35]
    \mpsWire{0}{0}{0}{5}
    \foreach \t in {1,...,4}{\gridLine{0}{\t}{2}{\t}}
    \mpsA{0}{4}{colMPS}
    \mpsB{0}{3}{colMPS}
    \mpsC{0}{1}{2}{colMPS}
    \gridLine{1}{2}{1}{4}
    \bCircle{1}{3}{FcolU}
  \end{tikzpicture}=
  \begin{tikzpicture}[baseline={([yshift=-0.6ex]current bounding box.center)},scale=0.35]
    \mpsWire{0}{0}{0}{5}
    \foreach \t in {1,...,4}{\gridLine{0}{\t}{1}{\t}}
    \mpsC{0}{3}{4}{colMPS}
    \mpsB{0}{2}{colMPS}
    \mpsA{0}{1}{colMPS}
  \end{tikzpicture},\qquad
  \begin{tikzpicture}[baseline={([yshift=-0.6ex]current bounding box.center)},scale=0.35]
    \mpsWire{0}{0}{0}{4}
    \foreach \t in {1,...,3}{\gridLine{0}{\t}{2}{\t}}
    \mpsB{0}{3}{colMPS}
    \mpsC{0}{1}{2}{colMPS}
    \gridLine{1}{2}{1}{4}
    \ME{1}{4}
    \mpsBvec{0}{4}
    \bCircle{1}{3}{FcolU}
  \end{tikzpicture}=
  \begin{tikzpicture}[baseline={([yshift=-0.6ex]current bounding box.center)},scale=0.35]
    \mpsWire{0}{0}{0}{4}
    \foreach \t in {1,...,3}{\gridLine{0}{\t}{1}{\t}}
    \mpsBvec{0}{4}
    \mpsA{0}{3}{colMPS}
    \mpsB{0}{2}{colMPS}
    \mpsA{0}{1}{colMPS}
  \end{tikzpicture},\qquad
  \begin{tikzpicture}[baseline={([yshift=-0.6ex]current bounding box.center)},scale=0.35]
    \mpsWire{0}{0}{0}{3}
    \foreach \t in {1,...,2}{\gridLine{0}{\t}{2}{\t}}
    \mpsC{0}{1}{2}{colMPS}
    \gridLine{1}{2}{1}{3}
    \ME{1}{3}
    \mpsBvec{0}{3}
  \end{tikzpicture}=
  \begin{tikzpicture}[baseline={([yshift=-0.6ex]current bounding box.center)},scale=0.35]
    \mpsWire{0}{0}{0}{3}
    \foreach \t in {1,...,2}{\gridLine{0}{\t}{1}{\t}}
    \mpsBvec{0}{3}
    \mpsB{0}{2}{colMPS}
    \mpsA{0}{1}{colMPS}
  \end{tikzpicture},
\end{gather}
where we have also introduced an additional tensor that acts on two physical
degrees of freedom to be able to formulate the algebraic conditions.  Using
this set of relations we can see that the two layers of $\tilde{\mathbb{W}}$
and $\tilde{\mathbb{W}}^{\prime}$ map between $\bra{L}$ and $\bra{L^{\prime}}$
and so we have
\begin{equation}
  \bra{L}\tilde{\mathbb{W}}=\Lambda \bra{L},\qquad 
  \bra{L^{\prime}}\tilde{\mathbb{W}}^{\prime}=\Lambda \bra{L^{\prime}}.
\end{equation}
Note that the choice of $\lambda$ in eq.~\eqref{eq:bottomRelations} is somewhat
arbitrary as it could be absorbed in one of the two bottom tensors, but for now
we keep it general to be able to choose a convenient normalisation. 

Similarly, we can take the following ansatz for the right fixed points
\begin{equation}
  \ket{R}=\begin{tikzpicture}[baseline={([yshift=-0.6ex]current bounding box.center)},scale=0.35]   \mpsWire{0}{0}{0}{7}
    \vLine{0}{0}{-1}{0}
    \foreach \t in {1,...,6}{\gridLine{0}{\t}{-1}{\t}}
    \foreach \t in {1,3,...,6}{\mpsA{0}{\t}{colMPS}}
    \foreach \t in {2,4,...,6}{\mpsB{0}{\t}{colMPS}}
    \mpsBvecW{0}{0}{colMPS}
    \mpsBvec{0}{7}
  \end{tikzpicture},\qquad
  \ket{R^{\prime}}=
  \begin{tikzpicture}[baseline={([yshift=-0.6ex]current bounding box.center)},scale=0.35]
    \mpsWire{0}{0}{0}{7}
    \vLine{0}{0}{-1}{0}
    \foreach \t in {1,...,6}{\gridLine{0}{\t}{-1}{\t}}
    \foreach \t in {1,3,...,6}{\mpsB{0}{\t}{colMPS}}
    \foreach \t in {2,4,...,6}{\mpsA{0}{\t}{colMPS}}
    \mpsBvecV{0}{0}{colMPS}
    \mpsBvec{0}{7}
  \end{tikzpicture},
\end{equation}
and assuming the left-right-flipped version of
eqs.~(\ref{eq:bottomRelations},\ref{eq:bulkTopRelations}) we get
\begin{equation}
  \tilde{\mathbb{W}}\ket{R}=\Lambda\ket{R},\qquad
  \tilde{\mathbb{W}}^{\prime}\ket{R^{\prime}}=\Lambda\ket{R^{\prime}}.
\end{equation}

\subsection{Parametrisation of left and right tensors} 
The above set of algebraic relations is a sufficient (but not necessary) condition
for the existence of influence matrices with small bond dimension, and at the moment
there are very few known cases with finite-dimensional solutions to these relations.
Interestingly in our case these relations can be solved by a set of $12$-dimensional
matrices. In particular, both on the left and the right we have the same top tensor
\begin{equation}
\begin{tikzpicture}[baseline={([yshift=-0.6ex]current bounding box.center)},scale=0.35]
  \mpsWire{0}{0}{0}{0.75}
  \mpsBvec{0}{0.75}
\end{tikzpicture}=\bra{t},
\end{equation}
which is a $12$-dimensional (row) vector. The bottom tensors on the left and right
are different (which is not surprising, as the matrices $W_{s,b}$, $V_{s,b}$ are not
symmetric), and we denote them by
\begin{equation}
  \begin{tikzpicture}[baseline={([yshift=-0.6ex]current bounding box.center)},scale=0.35]
    \mpsWire{0}{0}{0}{0.75}
    \vLine{0}{0}{0.75}{0}
    \mpsBvecW{0}{0}{colMPS}
    \node[anchor=west,inner sep=1pt] at (0.75,0) {\scalebox{0.8}{$x$}};
  \end{tikzpicture}=\ket*{b^{(L,1)}_{x}},\qquad
  \begin{tikzpicture}[baseline={([yshift=-0.6ex]current bounding box.center)},scale=0.35]
    \mpsWire{0}{0}{0}{0.75}
    \vLine{0}{0}{0.75}{0}
    \mpsBvecV{0}{0}{colMPS}
    \node[anchor=west,inner sep=1pt] at (0.75,0) {\scalebox{0.8}{$x$}};
  \end{tikzpicture}=\ket*{b^{(L,2)}_{x}},\\
  \begin{tikzpicture}[baseline={([yshift=-0.6ex]current bounding box.center)},scale=0.35]
    \mpsWire{0}{0}{0}{0.75}
    \vLine{0}{0}{-0.75}{0}
    \mpsBvecW{0}{0}{colMPS}
    \node[anchor=east,inner sep=1pt] at (-0.75,0) {\scalebox{0.8}{$x$}};
  \end{tikzpicture}=\ket*{b^{(R,1)}_{x}},\qquad
  \begin{tikzpicture}[baseline={([yshift=-0.6ex]current bounding box.center)},scale=0.35]
    \mpsWire{0}{0}{0}{0.75}
    \vLine{0}{0}{-0.75}{0}
    \mpsBvecV{0}{0}{colMPS}
    \node[anchor=east,inner sep=1pt] at (-0.75,0) {\scalebox{0.8}{$x$}};
  \end{tikzpicture}=\ket*{b^{(R,2)}_{x}},\qquad
\end{equation}
with $x\in\{0,1,2,3\}$ being a component in the auxiliary space of the stationary MPO. 
The bulk tensors are $12\times 12$ matrices and are the same on the left and right, just
evaluated at different parameters
\begin{equation}
  \begin{aligned}
    \begin{tikzpicture}[baseline={([yshift=-0.6ex]current bounding box.center)},scale=0.35]
      \mpsWire{0}{-0.625}{0}{0.625}
      \gridLine{0}{0}{0.625}{0}
      \mpsA{0}{0}{colMPS}
      \node[anchor=west,inner sep=1pt] at (0.625,0) {\scalebox{0.8}{$sb$}};
    \end{tikzpicture} &= 
    A_{sb}(\alpha_L),\qquad&
    \begin{tikzpicture}[baseline={([yshift=-0.6ex]current bounding box.center)},scale=0.35]
      \mpsWire{0}{-0.625}{0}{0.625}
      \gridLine{0}{0}{0.625}{0}
      \mpsB{0}{0}{colMPS}
      \node[anchor=west,inner sep=1pt] at (0.625,0) {\scalebox{0.8}{$sb$}};
    \end{tikzpicture} &= 
    B_{sb}(\alpha_L),\qquad&
    \begin{tikzpicture}[baseline={([yshift=-0.6ex]current bounding box.center)},scale=0.35]
      \mpsWire{0}{-0.625}{0}{1.625}
      \gridLine{0}{0}{0.625}{0}
      \gridLine{0}{1}{0.625}{1}
      \mpsC{0}{0}{1}{colMPS}
      \node[anchor=west,inner sep=1pt] at (0.625,1) {\scalebox{0.8}{$s_1b_1$}};
      \node[anchor=west,inner sep=1pt] at (0.625,0) {\scalebox{0.8}{$s_2b_2$}};
    \end{tikzpicture} &= 
    C_{s_1b_1s_2b_2}(\alpha_L),\\
    \begin{tikzpicture}[baseline={([yshift=-0.6ex]current bounding box.center)},scale=0.35]
      \mpsWire{0}{-0.625}{0}{0.625}
      \gridLine{0}{0}{-0.625}{0}
      \mpsA{0}{0}{colMPS}
      \node[anchor=east,inner sep=1pt] at (-0.625,0) {\scalebox{0.8}{$sb$}};
    \end{tikzpicture} &= 
    A_{sb}(\alpha_R),&
    \begin{tikzpicture}[baseline={([yshift=-0.6ex]current bounding box.center)},scale=0.35]
      \mpsWire{0}{-0.625}{0}{0.625}
      \gridLine{0}{0}{-0.625}{0}
      \mpsB{0}{0}{colMPS}
      \node[anchor=east,inner sep=1pt] at (-0.625,0) {\scalebox{0.8}{$sb$}};
    \end{tikzpicture} &= B_{sb}(\alpha_R),&
    \begin{tikzpicture}[baseline={([yshift=-0.6ex]current bounding box.center)},scale=0.35]
      \mpsWire{0}{-0.625}{0}{1.625}
      \gridLine{0}{0}{-0.625}{0}
      \gridLine{0}{1}{-0.625}{1}
      \mpsC{0}{0}{1}{colMPS}
      \node[anchor=east,inner sep=1pt] at (-0.625,1) {\scalebox{0.8}{$s_1b_1$}};
      \node[anchor=east,inner sep=1pt] at (-0.625,0) {\scalebox{0.8}{$s_2b_2$}};
    \end{tikzpicture} &= 
    C_{s_1b_1s_2b_2}(\alpha_R).
  \end{aligned}
\end{equation}
Explicit forms of the boundary tensors and the full parametrisation of bulk matrices are given in appendix~\ref{sec:ExplicitTensors}. The parameters $\alpha_L$ and $\alpha_R$ can be obtained from $\omega$, $\xi$ and $\Lambda$ as
\begin{equation}\label{eq:paramsAlpha}
  \alpha_L=
  \frac{\omega \Lambda(\Lambda-1)+\xi(\xi-\omega^2)}{\Lambda(\xi^2-\omega+\Lambda\omega)}
  ,\qquad
  \alpha_L\xleftrightarrow{\omega\leftrightarrow\xi}\alpha_R,
\end{equation}
where we note that $\Lambda$ is invariant under the swap of $\xi$ and $\omega$.

Given that the set of algebraic relations is finite and they involve finite-dimensional spaces, it is straightforward to verify (at least using computer algebra systems) that the objects reported in appendix~\ref{sec:ExplicitTensors} indeed solve the conditions and therefore constitute exact influence matrices. It can also be checked that for finite times these influence matrices cannot be further suppressed, suggesting that the bond dimension is indeed optimal. 

\subsection{Convenient reparametrisation of stationary states and Bethe ansatz}
Before moving to multi-time correlation functions it makes sense to discuss the above parametrisation of fixed points. First we observe that the transformation $(\xi,\omega) \to (\alpha_L,\alpha_R)$ maps $(0,\infty)\cross (0,\infty)$ to $(0,1) \cross (0,1)$. Moreover, in this range the mapping can be inverted to get
\begin{equation} \label{eq:xiomegaOfAlpha}
  \xi=\frac{1-\alpha_L}{\alpha_L^{\frac{4}{3}}\alpha_R^{\frac{2}{3}}},\qquad
  \omega=\frac{1-\alpha_R}{\alpha_R^{\frac{4}{3}}\alpha_L^{\frac{2}{3}}},\qquad
  \Lambda=\frac{1}{\alpha_L\alpha_R},
\end{equation}
and so we can equivalently rewrite stationary states $\rho$ in terms of $0<\alpha_L,\alpha_R<1$.
For convenience, let us define $0<\vartheta_{\pm}<1$
\begin{equation}
  \vartheta_{+}=1-\alpha_L,\qquad \vartheta_{-}=1-\alpha_R,
\end{equation}
and 
\begin{equation}
  \eta_{\pm}=\frac{1-\vartheta_{\pm}}{\vartheta_{\pm}}.
\end{equation}
Using these definitions together with the expressions of $\xi$ and $\omega$ in
terms of chemical potentials $\mu_{\pm}$ given in eq.~\eqref{eq:defFugacities}
we can rewrite eq.~\eqref{eq:xiomegaOfAlpha} as
\begin{equation}\label{eq:TBAthermal}
  \log\eta_{\nu}=\mu_{\nu} + 
  \smashoperator{\sum_{\nu^{\prime}\in\{+,-\}}} T_{\nu\nu^{\prime}}
  \log\left[1+\eta_{\nu^{\prime}}^{-1}\right],
\end{equation}
where $\nu\in\{+,-\}$ and we have introduced 
\begin{equation}
  T_{\nu,\nu^{\prime}}=
  \begin{cases}
    \frac{1}{3},& \nu=\nu^{\prime},\\
    \frac{2}{3},& \nu\neq\nu^{\prime}.
  \end{cases}
\end{equation}
Moreover, the free-energy density is
\begin{equation}\label{eq:TBAfreeenergy}
  \frac{2\log Z}{L}=
  \sum_{\nu}\log \left(1+\eta_{\nu}^{-1}\right).
\end{equation}
eqs.\ (\ref{eq:TBAthermal}-\ref{eq:TBAfreeenergy}) are reminiscent of
\emph{thermodynamic Bethe ansatz} (TBA)~\cite{takahashi1999thermodynamics}.
Indeed, if we interpret $\vartheta_{\nu}$ as the filling function of the mode
$\nu$ and $T_{\nu,\nu^{\prime}}$ as the scattering kernel, then
eq.~\eqref{eq:TBAthermal} is the thermal TBA equation for the Gibbs state in
eq.~\eqref{eq:defGibbs} in a model with two modes (labeled by $\nu\in\{+,-\}$).
eq.~\eqref{eq:TBAfreeenergy} is the corresponding equation for the extensive part 
of the partition function.  Similar expressions have been obtained for the Rule 54 cellular automaton~\cite{friedman2019integrable}.

Analogously one can show that traces of powers of the Gibbs state can be rewritten
as
\begin{equation}
  \frac{2\log \tr[\rho^n]}{L}=\sum_{\nu}\log\left(\frac{1+\eta_{n,\nu}^{-1}}{
    \left(1+\eta^{-1}\right)^n}\right),
\end{equation}
with $\eta_{n,\nu}$ satisfying
\begin{equation}
  \log\eta_{n,\nu}=n\mu_{\nu} + 
  \smashoperator{\sum_{\nu^{\prime}\in\{+,-\}}} T_{\nu\nu^{\prime}}
  \log\left[1+\eta_{n,\nu^{\prime}}^{-1}\right].
\end{equation}
These again take the expected TBA form~\cite{alba2017quench}.

\begin{figure}
  \centering
  \includegraphics[width=0.45\textwidth]{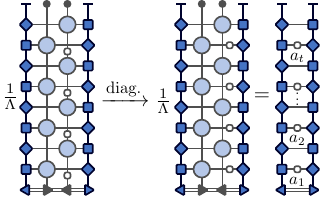}
  \caption{\label{fig:multiTimeD} Multi-time correlation functions at the same position (left). If the observables are diagonal in the computational basis, they can be commuted through the connections of lines and move them to the side (middle), and then use the fixed-point property of influence matrices to express it as a matrix element between $\bra{L}$ and $\ket{R}$ (right).}
\end{figure}

\section{Multi-time correlations of one-site diagonal observables}\label{sec:MultiTimeDiag}
Influence matrices encode all the information about the effective bath produced by a thermodynamically large system on its finite parts. Therefore, if one is interested in quantities constrained to a finite subsystem, one can use the influence matrices to reduce the calculation to a finite size. The quantity that exemplifies this feature in the most efficient way are multi-time correlation functions of local one-site observables at the same position,
\begin{equation}
  M_{a_1a_2\ldots a_{t}}=
  \frac{\tr[
    \left(\mathbb{U}^{\mathrm{o}}\mathbb{U}^{\mathrm{e}}\right)^{\dagger\, t}
    \rho a_1 \mathbb{U}^{\mathrm{o}} \mathbb{U}^{\mathrm{e}} a_2 
    \mathbb{U}^{\mathrm{o}}\mathbb{U}^{\mathrm{e}} a_3 \cdots ]}{\tr \rho}
  =\frac{1}{\Lambda}\mel{L}{\tilde{\mathbb{W}}[a_1,a_2,\ldots,a_{t}]}{R},
\end{equation}
where $\tilde{\mathbb{W}}[a_1,a_2,\ldots,a_{t-1}]$ is a multi-observable
generalisation of $\tilde{\mathbb{W}}[a,b]$ as shown on the left of
figure~\ref{fig:multiTimeD}. We have thus reduced the calculation to a subsystem
of length $2$. We can do better if the observables $a_j$ are diagonal in the
computational basis. In this case they can be moved around the crossing of
lines and the calculation reduces to a simple matrix element as shown on the
right of figure~\ref{fig:multiTimeD}. 

This is a big simplification, as now we only need to deal with matrices acting
on the tensor product of two auxiliary spaces,
\begin{equation}
  A_{s,b}\otimes B_{s,b},\quad \text{and}\quad
  B_{s,b}\otimes A_{s,b}.
\end{equation}
However, these are still 144-dimensional, and the information can be further
compressed. To see that, we introduce a projector from the full 12-dimensional
auxiliary space to a 6-dimensional subspace
\begin{equation}
  P= \begin{bmatrix}
    1 & 0&0&0\\
    0 & 0&0&1
  \end{bmatrix}\otimes \unit_3,
\end{equation}
where $\unit_3$ is a $3\times 3$ identity matrix. The projector acts trivially on
the top and bottom boundary vectors,
\begin{equation}
  \bra{t} P P^T=\bra{t},
\end{equation}
and
\begin{equation}
  P P^T\ket*{b_x^{(L/R,1/2)}}=\ket*{b_x^{(L/R,1/2)}}.
\end{equation}
Then one can straightforwardly verify that also the following holds,
\begin{equation}
  \left(P^T \otimes P^T\right) A_{s,b}\otimes B_{s,b}
  =\left(P^T \otimes P^T\right) A_{s,b}\otimes B_{s,b} \left(P P^T\otimes P P^T\right),
\end{equation}
which implies that in the expression for $M_{a_1\ldots a_t}$ we can make the replacement
\begin{equation}
    A_{s,b}\mapsto \tilde{A}_{s,b}=P^T A_{s,b} P,\qquad
    B_{s,b}\mapsto \tilde{B}_{s,b}=P^T B_{s,b} P,
\end{equation}
(analogously we define $\ket*{\tilde{b}}$ and $\bra{\tilde{t}}$), and so we
have reduced the required bond dimension to~$6$.

An additional reduction comes from an observation that
$\tilde{A}_{s,b}\otimes \tilde{B}_{s,b}$ and $\tilde{B}_{s,b}\otimes \tilde{A}_{s,b}$
are block diagonal with respect to the projectors 
\begin{equation}
  Q_2 = \left(\begin{bmatrix} 0 & 0 \\ 0 & 1 \end{bmatrix} \otimes \unit_3\right)^{\otimes 2},
    \qquad 
    Q_1 = \unit_{36}-Q_2,
\end{equation}
i.e.\ we have
\begin{equation}
  Q_j \tilde{A}_{s,b}\otimes \tilde{B}_{s,b} =
  Q_j \tilde{A}_{s,b}\otimes \tilde{B}_{s,b} Q_j,\qquad
  Q_j \tilde{B}_{s,b}\otimes \tilde{A}_{s,b} =
  Q_j \tilde{B}_{s,b}\otimes \tilde{A}_{s,b} Q_j,
\end{equation}
and furthermore, the bottom state lives in the sector given by $Q_1$,
\begin{equation}
  Q_2 \sum_{x=0}^3 P^T \ket*{b_x^{(L,1/2)}}\otimes P^T\ket*{b_x^{(R,2/1)}}=0,
\end{equation} 
meaning that the relevant auxiliary space necessary to evaluate $M_{a_1\ldots
a_t}$ is immediately reduced to be $27$-dimensional (down from 36). This
simplification is not optimal, as finite-time numerics indicates that
the MPO for $M_{a_1\ldots a_t}$ can be further reduced, but we have not
attempted to do it at this point.  The main motivation behind finding  $Q_1$
and $Q_2$ was to get rid of the block diagonal structure in the auxiliary
space, which can (as we argue below) introduce degeneracies when evaluating the
multi-time correlation function.

\subsection{Two-time autocorrelation function}
To demonstrate the effectiveness of these manipulations, let us consider a
two-time autocorrelation function, that is $M_{a_1 a_2\ldots a_{t+1}}$ evaluated
for $a_1=a$, $a_{t+1}=b$ and $a_2=\cdots=a_{t}=\unit$. Explicitly,
\begin{equation}\label{eq:defCabtD}
  C_{a,b}(t)=
  \bra*{\tilde{T}}{\tilde{M}[b] \tilde{M}^{t-1} \tilde{M}[a]}\ket*{\tilde{B}},
\end{equation}
where we have introduced reduced boundary vectors on two legs as
\begin{equation}
  \bra*{\tilde{T}} = \bra{t} P \otimes \bra{t} P,\qquad
  \ket*{\tilde{B}}
  =\sum_{x=0}^3 
  P^T\ket*{b_x^{(L,2)}}\otimes P^T\ket*{b_x^{(R,1)}},
\end{equation}
and $\tilde{M}[o]$, $\tilde{M}=\tilde{M}[\unit]$ are the bulk transfer matrices
\begin{equation}
    \tilde{M}[a]=
    \left(\sum_{s,b}\tilde{A}_{s,b}(\alpha_L)\otimes \tilde{B}_{s,b}(\alpha_R)\right)
    \left(
    \sum_{s,b}
    \mel{s}{a}{s}
    \tilde{B}_{s,b}(\alpha_L)\otimes \tilde{A}_{s,b}(\alpha_R)\right).
\end{equation}

As the matrix $\tilde{M}$ is finite-dimensional it suffices to diagonalise it to
characterise the large-$t$ asymptotics of $C_{a,b}(t)$. We first note that the algebraic
relations imply that the top and bottom boundary are left and right eigenvectors
of $\tilde{M}$,
\begin{equation}
  \bra*{\tilde{T}}\tilde{M}=\bra*{\tilde{T}},\qquad
  \tilde{M}\ket*{\tilde{B}}=\ket*{\tilde{B}}.
\end{equation}
This immediately implies that correlation functions of observables that have $\unit$
as a component do not decay in time, which suggests that it makes sense to restrict
the discussion to observables whose stationary expectation value is $0$,
\begin{equation}
  a\to a-\unit \expval{a},
\end{equation}
or equivalently, to consider the \emph{connected} part of the correlation
function $C_{a,b}(t)$. 

\begin{figure}
  \centering
  \includegraphics[width=0.65\textwidth]{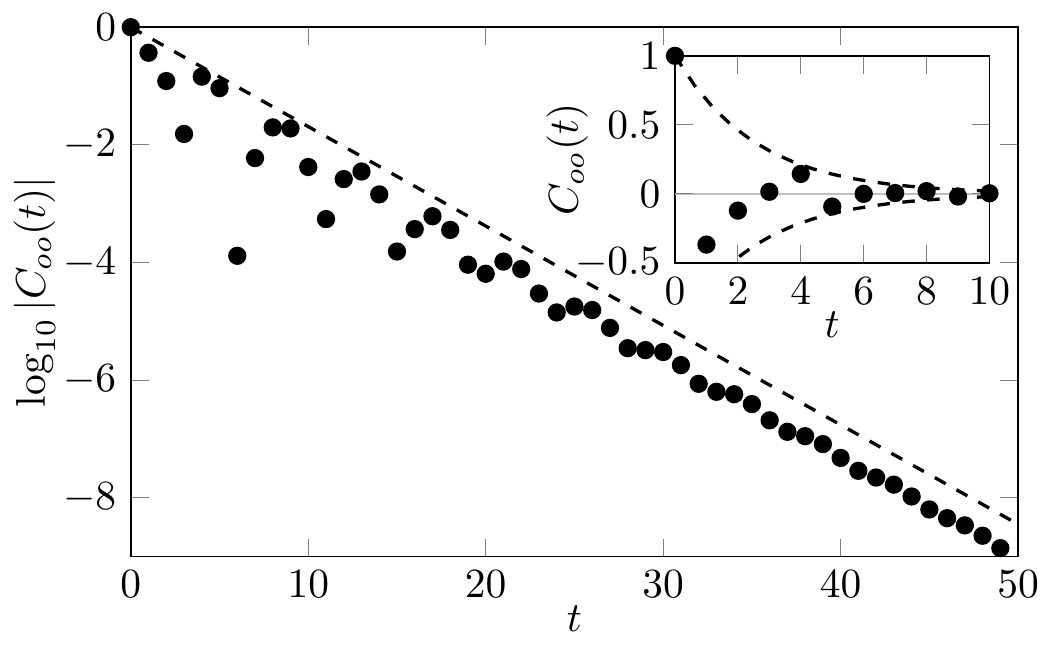}
  \caption{\label{fig:corrDecay} Correlation function $C_{oo}(t)$ for $(\alpha_L,\alpha_R)=(0.4,0.7)$ and observable $o$ as defined in eq.~\eqref{eq:defOtraceless}. The asymptotic decay given by $\lambda_1\approx -0.677569$ is reported with dashed lines.}
\end{figure}

Explicitly diagonalising the full matrix $\tilde{M}$ we observe that the
eigenvalue $1$ is two-fold degenerate, which could in principle imply that even
some \emph{connected} correlation functions tend to a constant. However, this
is not the case, as the bottom boundary state $\ket*{\tilde{B}}$ restricts us
to one of the two invariant subspaces as discussed above. In other words, for
correlation functions the relevant transfer matrix is not the full $\tilde{M}$,
but its block preserved by the projector $Q_1$,
\begin{equation}
  \tilde{M}\mapsto Q_1\tilde{M} Q_1,
\end{equation}
and with an explicit calculation one obtains
\begin{equation}
  \mathrm{Spect}(Q_1\tilde{M} Q_1)=\{1,\lambda_1,\lambda_2,\lambda_3,\lambda_4,0\},
\end{equation}
where all the non-zero eigenvalues are isolated and $\lambda_j$ are the four roots 
of the quartic equation
\begin{equation}\label{eq:defCorrPoly}
  \begin{aligned}
    \lambda^4&+\lambda^3+\Big[1-2(1-\alpha_L)(1-\alpha_R)\Big]\lambda^2\\
    &+\Big[1-\alpha_L+1-\alpha_R-3(1-\alpha_L)(1-\alpha_R)\Big]\lambda
    +(1-\alpha_L)(1-\alpha_R)\alpha_L\alpha_R=0.
  \end{aligned}
\end{equation}
The asymptotic decay of correlation functions is therefore expected to be
governed by the largest (in magnitude) solution to eq.~\eqref{eq:defCorrPoly}.
In figure~\ref{fig:corrDecay} we demonstrate that this is indeed the case for 
\begin{equation}\label{eq:defOtraceless}
  \mkern-20mu
  o=\frac{1}{2\sqrt{3-\alpha_L\alpha_R}}
  \left((4-\alpha_L\alpha_R)\sigma_z-(2-\alpha_L\alpha_R)\unit\right),
  \mkern-20mu
\end{equation}
which is up to a prefactor the only one-site diagonal observable whose expectation
value vanishes. 

\section{Conclusions}\label{sec:conclusions}
In this work we have studied the finite-subsystem dynamics of Rule 201, i.e.\ a
classical reversible cellular automaton that implements the PXP constraint. As
long as the subsystem is kept finite, all its dynamical properties can be
exactly described using finite-size evolution operators with appropriate
boundary conditions set by the influence matrices. For the class of Gibbs
states, we have shown that the influence matrices can be rewritten as MPOs of
bond dimension 12. We have shown that these MPOs are naturally parametrised by
solutions to TBA-like equations, even though the Bethe ansatz for this model
has not yet been developed. Moreover, we have shown that using our influence
matrices, one can completely characterise the decay of one-site autocorrelation
functions.

Our results open a number of interesting questions for future research. An
immediate one is to find a Bethe-ansatz description of the model and provide 
an ab-initio derivation of the TBA equations identified here. We are currently
working on a coordinate-Bethe ansatz approach~\cite{edge2026integrability}.

Another interesting question concerns the physics contained in these influence
matrices. In this work we have shown that the information can be greatly compressed
if we only care about one-site observables. A natural direction is to ask
what are the physical quantities that require the full $12$-dimensional MPO.

Finally, an important question is to generalise this approach to quench
problems. More specifically, to ascertain whether there exist initial states
that generate finite-dimensional influence matrices compatible to the ones
found here. This would, e.g., allow us to gain an analytical handle on the
growth of entanglement entropies, and a deeper understanding of the dynamics
of quantum information in interacting integrable
models~\cite{bertini2022growth,bertini2023nonequilibrium,bertini2024dynamics}.

\acknowledgments
I would like to thank Bruno Bertini, Adam Insall, and Thomas Edge for insightful discussions and collaboration on related topics, and Bruno Bertini for the careful reading of the manuscript. 

\vspace{0.5cm}
\emph{Note added: while preparing this manuscript we became aware of the Ref.~\cite{yang2026solving}. Where overlapping, our results seem compatible.}

\appendix
\addtocontents{toc}{\protect\setcounter{tocdepth}{1}}
\section{Fixed-point MPS}\label{sec:ExplicitTensors}
\subsection{Main bulk tensors}
The physical-space components of the bulk tensors are $12\times 12$ matrices,
acting on an auxiliary space which we will for compactness represent as a
tensor product of a $3$ and $4$-dimensional vector space. We introduce
$\mathrm{e}_{i,j}$ with $1\le i,j\le 4$ to denote basis elements in the space
of $4\times 4$ matrices
\begin{equation}
  \mathrm{e}_{i,j}=\ketbra{i}{j},
\end{equation}
and $\unit_3$ to be a $3\times 3$ identity matrix. Matrices $B_{sb}$ can then
be compactly represented as
\begin{equation}
  B_{00}=
    \mathrm{e}_{11} \otimes
    \unit_3,\qquad
    B_{01}=
    \mathrm{e}_{22} \otimes
    \unit_3,\qquad
    B_{10}=\mathrm{e}_{33} \otimes
    \unit_3,\qquad
    B_{11}=
    \mathrm{e}_{44} \otimes
    \unit_3.
\end{equation}
Here we note that $B_{10}$ is related to $B_{01}$ through a swain the $4$-dimensional space, 
\begin{equation}
  B_{10}=S B_{01} S^{-1},\qquad
  S=\begin{bmatrix}
    1 &   &   & \\
      & 0 & 1 & \\
      & 1 & 0 & \\
      &   &   & 1
  \end{bmatrix}\otimes \unit_3.
\end{equation}
In fact, this is a symmetry of \emph{all} bulk tensors 
\begin{equation}\label{eq:TensorSymmetry}
  A_{sb}=S A_{b s} S^{-1},\qquad
  B_{sb}=S B_{b s} S^{-1},
\end{equation}
and therefore it suffices to report only $3$ out of $4$ physical space components of tensors $A_{sb}$,
\begin{equation}
  \mkern-10mu
  \begin{aligned}
    A_{00}&=\mathrm{e}_{11}\otimes
    \begin{bmatrix} 
      0 & 0 & 0 \\ 
      0 & 0 & 1 \\ 
      0 & 0 & 0 
    \end{bmatrix}
    +
    \mathrm{e}_{14}\otimes
    \begin{bmatrix} 
      0 & 1 & 0 \\ 
      0 & 0 & 0 \\ 
      1 & 0 & 1 
    \end{bmatrix}
    +
    \left(\mathrm{e}_{23} +\mathrm{e}_{32}\right)\otimes
    \begin{bmatrix} 
      0 & 0 & 0 \\ 
      0 & 0 & 1\\
      \alpha & 0 & 0
    \end{bmatrix}
    +\mathrm{e}_{41} \otimes 
    \begin{bmatrix} 
      0 & \alpha & 0 \\ 
      1-\alpha & 1-\alpha & 0 \\ 
      \alpha & 0 & 0 
    \end{bmatrix},\\
    A_{01}&=\mathrm{e}_{11}\otimes
    \begin{bmatrix} 
      0 & 0 & 0 \\ 
      0 & 0 & 1 \\ 
      0 & 0 & 0 
    \end{bmatrix}
    +
    \mathrm{e}_{13}\otimes
    \begin{bmatrix} 
      \alpha & 0 & 0 \\ 
      0 & 0 & 0 \\ 
      0 & 0 & 1 
    \end{bmatrix}
    +
    \mathrm{e}_{22} \otimes
    \begin{bmatrix} 
      0 & 1 & 0 \\ 
      0 & 0 & 0 \\ 
      0 & 0 & 0
    \end{bmatrix}
    +\mathrm{e}_{24} \otimes
    \begin{bmatrix} 
      0 & 0 & 0 \\ 
      1 & 0 & 0 \\
      0 & 1 & 0
    \end{bmatrix}\\
    &+\mathrm{e}_{31} \otimes 
    \begin{bmatrix} 
      0 & 0 & 0 \\
      1 & 0 & 0 \\
      0 & \alpha & 0 
    \end{bmatrix}
    + \mathrm{e}_{42} \otimes
    \begin{bmatrix}
      \alpha & 0 & 0 \\
      0 & 0 & 1-\alpha \\
      0 & 0 & \alpha
    \end{bmatrix},\\
    A_{11}&=\mathrm{e}_{11}\otimes
    \begin{bmatrix} 
      0 & \alpha & 0 \\ 
      0 & 0 & 1 \\ 
      1 & 1-\alpha & 0 
    \end{bmatrix}
    +
    \left(\mathrm{e}_{22}+\mathrm{e}_{33}\right)\otimes
    \begin{bmatrix} 
      0 & 1 & 0 \\ 
      0 & 0 & 1 \\ 
      \alpha & 0 & 0
    \end{bmatrix}
    +\mathrm{e}_{44} \otimes
    \begin{bmatrix} 
      0 & 1 & 0 \\ 
      1-\alpha & 0 & 1 \\
      \alpha & 0 & 0
    \end{bmatrix}.
  \end{aligned}
  \mkern-40mu
\end{equation}
We report the rest of the bulk tensors in appendix~\ref{sec:lastTensors}.

\subsection{Boundary tensors}\label{sec:boundaryTensors}
We now proceed to boundary tensors. Note that with the choice below we have fixed
the parameter $\lambda$ in eq.~\eqref{eq:bottomRelations} to be equal to 
\begin{equation}
  \lambda=\frac{1}{\alpha_R},
\end{equation}
and we have fixed the normalisation so that
\begin{equation}
    1=\sum_{x=0}^3 \braket*{t}{b_x^{(L,1)}} \braket*{t}{b_x^{(R,2)}}
    =\sum_{x=0}^3 \braket*{t}{b_x^{(L,2)}} \braket*{t}{b_x^{(R,1)}}.
\end{equation}
This property, together with
\begin{equation}
  \sum_{s,b} \bra{t}A_{s,b}(\alpha_L) \otimes \bra{t}B_{s,b}(\alpha_R)
  =\sum_{s,b} \bra{t}B_{s,b}(\alpha_L) \otimes \bra{t}A_{s,b}(\alpha_R)=
  \bra{t}\otimes\bra{t},
\end{equation}
implies that the left and right influence matrices are normalised as
\begin{equation}
  \braket*{L}{R}=1.
\end{equation}

To express the boundary tensors we use the fact that they are nontrivial on an
effectively $6$-dimensional space as described in section~\ref{sec:MultiTimeDiag}.
Explicitly, we have
\begin{equation}
  \bra{t} = \bra{\tilde{t}} P^T,\qquad
  \ket*{b_{x}^{L/R,j}} = P \ket*{\tilde{b}_{x}^{L/R,j}}
\end{equation}
with
\begin{equation}
  P=\begin{bmatrix}
    1 & 0 & 0 & 0 \\ 0 & 0 & 0 & 1
  \end{bmatrix}\otimes \unit_3.
\end{equation}
The reduced top boundary vector can be then given as
\begin{equation}
  \bra{\tilde{t}} =
  \begin{bmatrix}
    1&1&1&1&1&1
  \end{bmatrix}.
\end{equation}
The bottom tensors act on the $4$-dim auxiliary space of the stationary state,
and their components are $12$-dimensional vectors in auxiliary spaces of
influence matrices and their reduced counterparts on the left are
\begin{equation}
  \begin{aligned}
    \ket*{\tilde{b}_{0}^{(L,1)}}&=
    \frac{\alpha_L^{\frac{4}{3}}\alpha_R^{\frac{2}{3}}}{1-\alpha_L}
    \begin{bmatrix}
      1 \\ 
      \frac{\alpha_R}{1-\alpha_R}\\
      \frac{1-\alpha_L}{\alpha_L(1-\alpha_R)}\\
      1 \\
      \frac{1-\alpha_L}{\alpha_L \alpha_R}\\
      \frac{1-\alpha_R}{\alpha_R}
    \end{bmatrix},&
    \ket*{\tilde{b}_{1}^{(L,1)}}&=
      \frac{\alpha_L^{\frac{2}{3}}\alpha_R^{\frac{4}{3}}}{1-\alpha_R}
    \begin{bmatrix}
      1 \\ \frac{1-\alpha_R}{\alpha_R(1-\alpha_L)} \\ \frac{\alpha_L}{1-\alpha_L} \\ 0 \\ 0 \\ 0
    \end{bmatrix},\\
    \ket*{\tilde{b}_{2}^{(L,1)}}&=
    \begin{bmatrix} 
      1 \\ \frac{\alpha_R}{1-\alpha_R} \\ \frac{\alpha_L}{1-\alpha_L} \\
      \frac{\alpha_L^2 \alpha_R^2}{(1-\alpha_L)(1-\alpha_R)} \\
      \frac{\alpha_L \alpha_R}{1-\alpha_R} \\
    \frac{\alpha_L^2\alpha_R}{1-\alpha_L}\end{bmatrix},&
    \ket*{\tilde{b}_{3}^{(L,1)}}&=
    \frac{\alpha_L^{\frac{2}{3}}(1-\alpha_R)}{\alpha_R^{\frac{2}{3}}(1-\alpha_L)}
    \begin{bmatrix}
      1 \\ \frac{\alpha_R}{1-\alpha_R} \\ \frac{1-\alpha_L}{\alpha_L(1-\alpha_R)} \\
      \frac{(1-\alpha_L)\alpha_R^2}{(1-\alpha_R)^2} \\ 
      \frac{(1-\alpha_L)^2 \alpha_R}{\alpha_L(1-\alpha_R)^2} \\ 
      \frac{(1-\alpha_L)\alpha_R}{1-\alpha_R}
    \end{bmatrix},\\
    \ket*{\tilde{b}_{0}^{(L,2)}}&=
    \frac{\alpha_L^{\frac{4}{3}}\alpha_R^{\frac{2}{3}}}{1-\alpha_R}
    \begin{bmatrix}
      1 \\ \frac{1-\alpha_R}{(1-\alpha_L)\alpha_R} \\ \frac{\alpha_L}{1-\alpha_L}\\
      1 \\ \frac{1-\alpha_L}{\alpha_L \alpha_R} \\ \frac{1-\alpha_R}{\alpha_R}\end{bmatrix},&
    \ket*{\tilde{b}_{1}^{(L,2)}}&=
    \frac{\alpha_L^2}{1-\alpha_L}
    \begin{bmatrix}
      1 \\ \frac{\alpha_R}{1-\alpha_R} \\ \frac{1-\alpha_L}{\alpha_L(1-\alpha_R)} \\
      0 \\ 0 \\ 0
    \end{bmatrix},\\
    \ket*{\tilde{b}_{2}^{(L,2)}}&=
    \frac{\alpha_L^{\frac{2}{3}}}{\alpha_R^{\frac{2}{3}}}
    \begin{bmatrix}
      1 \\ \frac{\alpha_R}{1-\alpha_R} \\ \frac{\alpha_L}{1-\alpha_L} \\
      \frac{\alpha_L^2\alpha_R^2}{(1-\alpha_L)(1-\alpha_R)} \\
      \frac{\alpha_L \alpha_R}{1-\alpha_R} \\
    \frac{\alpha_L^2\alpha_R}{1-\alpha_L}\end{bmatrix},&
    \ket*{\tilde{b}_{3}^{(L,2)}}&=\frac{1-\alpha_L}{1-\alpha_R}
    \begin{bmatrix}
      1 \\ \frac{1-\alpha_R}{(1-\alpha_L)\alpha_R} \\ \frac{\alpha_L}{1-\alpha_L} \\
      \frac{\alpha_L^2(1-\alpha_R)}{(1-\alpha_L)^2} \\
      \frac{\alpha_L(1-\alpha_R)}{(1-\alpha_L)\alpha_R} \\
      \frac{\alpha_L^2(1-\alpha_R)^2}{(1-\alpha_L)^2\alpha_R}
    \end{bmatrix},
  \end{aligned}
\end{equation}
while on the right we have
\begin{equation}
  \begin{aligned}
    \ket*{\tilde{b}_{0}^{(R,1)}}&=
    \frac{\alpha_R^{\frac{1}{3}}}{\alpha_L^{\frac{1}{3}}}\ket*{\tilde{b}_{2}^{(R,2)}}=
    \frac{(1-\alpha_L)(1-\alpha_R)\alpha_R^{\frac{1}{3}}}{\alpha_L^{\frac{1}{3}}(4-\alpha_L\alpha_R)}
    \begin{bmatrix} 0 \\ 1 \\ 0 \\ 0 \\ 0 \\ 0 \end{bmatrix},\\
    \ket*{\tilde{b}_{1}^{(R,1)}}&=
    \frac{\alpha_R^{\frac{1}{3}}}{\alpha_L^{\frac{1}{3}}}\ket*{\tilde{b}_{1}^{(R,2)}}=
    \frac{(1-\alpha_L)(1-\alpha_R)\alpha_R}{4-\alpha_L\alpha_R}
    \begin{bmatrix} 0 \\ 0 \\ 0 \\ 
      1 \\ \frac{1-\alpha_R}{\alpha_L \alpha_R} \\ \frac{1-\alpha_L}{\alpha_L}\end{bmatrix},\\
    \ket*{\tilde{b}_{2}^{(R,1)}}&=
    \frac{\alpha_R^{\frac{1}{3}}}{\alpha_L^{\frac{1}{3}}}\ket*{\tilde{b}_{0}^{(R,2)}}=
    \frac{(1-\alpha_L)(1-\alpha_R)\alpha_R^{\frac{2}{3}}}{\alpha_L^{\frac{2}{3}}(4-\alpha_L\alpha_R)}
    \begin{bmatrix} 0 \\ 0 \\ 1 \\ 0 \\ 0 \\ 0 \end{bmatrix},\\
    \ket*{\tilde{b}_{3}^{(R,1)}}&=
    \frac{\alpha_R^{\frac{1}{3}}}{\alpha_L^{\frac{1}{3}}}\ket*{\tilde{b}_{3}^{(R,2)}}=
    \frac{(1-\alpha_L)(1-\alpha_R)\alpha_R}{4-\alpha_L\alpha_R}
    \begin{bmatrix} 1 \\ 0 \\ 0 \\ 0 \\ 0 \\ 0 \end{bmatrix}.
  \end{aligned}
\end{equation}

\subsection{Two-site bulk tensors}\label{sec:lastTensors}
Finally, here we report the auxiliary two-physical-space matrices that
constitute the local relations, but do not appear in calculations of physically
motivated quantities. As before we make use of the symmetry under the swap of physical-space indices
\begin{equation}\label{eq:symmetryC}
  C_{s_1 b_1 s_2 b_2}= S C_{b_1 s_1 b_2 s_2} S^{-1},
\end{equation}
which allows us to explicitly give only $10$ out of $16$ auxiliary-space matrices, as the others are uniquely determined by~\eqref{eq:symmetryC}. Explicitly, for $s_1=b_1=0$ we get
\begin{equation}
  \mkern-20mu
  \begin{aligned}
    C_{0000}&=
    \mathrm{e}_{11}\otimes
    \begin{bmatrix}
      0&0&1-\alpha\\
      0&0&0\\
      0&0&\alpha
    \end{bmatrix}
    + \Big(\mathrm{e}_{14}+\alpha(\mathrm{e}_{22}+\mathrm{e}_{33})\Big)\otimes
    \begin{bmatrix}
      0&0&0\\
      0&\alpha&0\\
      0&0&0
    \end{bmatrix}+
    \mathrm{e}_{44}\otimes
    \begin{bmatrix}
      \alpha&0&0\\
      1-\alpha&0&1-\alpha\\
      0&0&\alpha
    \end{bmatrix},\\
    C_{0001}&=
    \mathrm{e}_{11}\otimes
    \begin{bmatrix}
      0&0&1-\alpha\\
      0&0&0\\
      0&0&\alpha
    \end{bmatrix}
    +\mathrm{e}_{13}\otimes
    \begin{bmatrix}
      0&0&0\\
      \alpha&0&0\\
      0&0&0
    \end{bmatrix}
    + \mathrm{e}_{33}\otimes
    \begin{bmatrix}
      0&0&0\\
      0&\alpha&0\\
      0&0&0
    \end{bmatrix}+
    \mathrm{e}_{43}\otimes
    \begin{bmatrix}
      0&0&\alpha\\
      0&0&1-\alpha\\
      0&0&0
    \end{bmatrix},\\
    C_{0011}&=
    \mathrm{e}_{11}\otimes
    \begin{bmatrix}
      0&0&1-\alpha\\
      0&\alpha&0\\
      0&0&\alpha
    \end{bmatrix}+
    \mathrm{e}_{41}\otimes
    \begin{bmatrix}
      \alpha&0&0\\
      1-\alpha&(1-\alpha)^2&0\\
      0&\alpha(1-\alpha)&0
    \end{bmatrix},
  \end{aligned}
  \mkern-60mu
\end{equation}
while the rest take the (slightly simpler) form
\begin{equation}
  \begin{aligned}
    C_{0100}&=
    \mathrm{e}_{13}\otimes
    \begin{bmatrix}
      0&0&0\\
      \alpha&0&0\\
      0&0&0
    \end{bmatrix}+
    \mathrm{e}_{33}\otimes
    \begin{bmatrix}
      0&0&0\\
      0&0&0\\
      0&0&\alpha
    \end{bmatrix},&
  C_{1100}&=\mathrm{e}_{11}\otimes
    \begin{bmatrix}
      \alpha&0&0\\
      0&\alpha&0\\
      1-\alpha&1-\alpha&0
    \end{bmatrix},\\
    C_{0111}&=\mathrm{e}_{12}
    \otimes
    \begin{bmatrix}
      0&0&0\\
      \alpha&0&0\\
      0&\alpha&0
    \end{bmatrix}
    +\mathrm{e}_{32}\otimes
    \begin{bmatrix}
      0&0&0 \\
      0&0&0 \\
      0&0&\alpha
    \end{bmatrix},&
    C_{1101}&=
    \mathrm{e}_{12}\otimes
    \begin{bmatrix}
      0&0&\alpha\\
      \alpha&0&0\\
      0&0&1-\alpha
    \end{bmatrix},\\
    C_{0110}&=
    \mathrm{e}_{11}\otimes
    \begin{bmatrix}
      0&0&0\\
      0&\alpha&0\\
      0&0&0
    \end{bmatrix}+
    \mathrm{e}_{31}\otimes
    \begin{bmatrix}
      0&0&0\\
      0&1-\alpha&0\\
      \alpha&0&0
    \end{bmatrix},&
    C_{1111}&=
    \mathrm{e}_{14}\otimes
    \begin{bmatrix}
      \alpha&0&0\\
      0&1&0\\
      1-\alpha&0&1
    \end{bmatrix},\\
    C_{0101}&=\mathrm{e}_{12}\otimes
    \begin{bmatrix}
      0&0&0 \\
      0&0&0 \\
      0&\alpha&0
    \end{bmatrix}
    +\mathrm{e}_{14}\otimes
    \begin{bmatrix}
      0&0&0\\
      0&1&0\\
      0&0&0
    \end{bmatrix}
    + \mathrm{e}_{34}\otimes
    \begin{bmatrix}
      0&0&0\\
      0&0&1\\
      \alpha&0&0
    \end{bmatrix}.\mkern-100mu&
  \end{aligned}
\end{equation}
\bibliographystyle{jsty3-author}
\bibliography{bibliography}
\end{document}